# An upper bound for the magnetic force gradient in graphite

David Martínez-Martín[1], Miriam Jaafar[1], Rubén Pérez[2], Julio Gómez-Herrero[1] and Agustina Asenjo[3].

[1]Dpto. Física de la Materia Condensada, UAM, 28049 Madrid, Spain

[2]Dpto. Física Teórica de la Materia Condensada, UAM, 28049 Madrid, Spain

[3]Instituto de Ciencia de Materiales de Madrid, CSIC, 28049 Madrid, Spain

Cervenka et al. have recently reported ferromagnetism along graphite steps. We present Magnetic Force microscopy (MFM) data showing that the signal along the steps is independent of an external magnetic field. Moreover, by combining Kelvin Probe Force Microscopy (KPFM) and MFM, we are able to separate the electrostatic and magnetic interactions along the steps obtaining an upper bound for the magnetic force gradient of ~16 $\mu$N/m, a figure six times lower than the lowest theoretical bound reported by Cervenka et al. Our experiments suggest absence of MFM signal in graphite at room temperature.

Ferromagnetism in carbon based materials [1-3] is a current hot topic with relevant implications in material science [4], nanotechnology [5], physics [6, 7] and even economy [8]. Ferromagnetism is usually associated with 3d and 4f orbitals but, in the case of carbon materials, it has been suggested that the presence of defects can be also at the origin of this phenomenon [9]. Most of the experiments reported so far are based on macroscopic techniques such as SQUID [1-3, 9]. STM has already shown that vacancies on the graphite surface exhibit a magnetic moment [7], but this does not necessarily implies ferromagnetism. Magnetic force microscopy [10, 11], a variation of atomic force microscopy (AFM) where magnetic probes are used, has the important advantage over SQUID of producing images where the spatial distribution of the magnetic signal can be resolved at the nanometer scale. In particular, AFM can easily detect small steps on a flat surface such as graphite. Since steps are always related to defects in the crystalline order, it is natural to assume that they should exhibit ferromagnetic character. Guided by this idea, Cervenka *et al.* [12] have recently reported MFM images showing contrast inversion along the steps when the magnetization of the tip is reversed. The implication is that the ferromagnetic domains are located in the grain boundaries, i.e. the spins along the surface steps should present a well defined orientation, namely up and down. The MFM data shown in reference [12] is complemented by SQUID measurements showing a clear ferromagnetic out-of-plane easy axis at 300 K (coercitive field of $H_c \approx$ 20 mT and a normalized remanence magnetization of about $M_{remanence}/M_{saturation} \approx 0.15$). As reported in the literature [13-16], MFM averaged quantities should be comparable to those obtained from SQUID and this argument also stands for the particular case of ferromagnetism in graphite as shown by Esquinazi *et al.* [17]. The SQUID data reported in ref. [2, 3, 12] imply that in the remanent state, the magnetization should present a distribution of about 60% in one direction and 40% in the other. Accordingly, MFM images taken in this magnetic state should exhibit a similar distribution of both brighter (associated with repulsive forces) and darker steps (associated with attractive forces) with respect to the terrace contrast. On the contrary, the MFM data reported in ref. [12], only show large size images where all the steps in the image are either dark or bright. In the supplementary information in ref. [12] steps showing different contrast are presented as a proof of the coexistence of different magnetic domains. However the contrast for both domains is well above the contrast observed in the terraces, suggesting a wrong interpretation. Moreover, the commercial cobalt-coated tips used in these experiments have a stray field ranging between 40mT and 60 mT [18-20]. As the tip scans the surface, a partial reversal magnetization of the sample is expected, leading to an increase of the number of dark contrast edges versus the number of bright edges. Assuming a magnetic origin for the tip-sample interaction, it would be extremely difficult to find large graphite areas with steps showing only bright contrast.

In this letter, we challenge those results and their interpretation using the same kind of sample and magnetic tips. We show that the contrast along the steps is independent of the external magnetic field and tip magnetization state. Furthermore, combining KPFM and MFM we demonstrate that most of the signal at the steps has an electrostatic origin. The magnetic force gradient associated with the remaining contrast, that we can easily detect with our more accurate phase-lock loop setup, is six times smaller than the lowest theoretical bound estimated in ref. [12]. Thus, we conclude that ferromagnetism is not present in graphite at room temperature.

We have carried out MFM experiments in Highly Oriented Pyrolytic Graphite (HOPG) samples of ZYH quality. HOPG samples were cleaved by an adhesive tape. The same samples were subsequently studied using two experimental set ups: first, an AFM in air ambient conditions with the capability to apply an out-of-plane external magnetic field $H_E$ between ±60 mT [21]. This field should be enough to reverse first the magnetization of the HOPG sample (assuming $H_c \approx 20$ mT [2, 3, 12]) and then the magnetization of the tip ($H_c = 45$ mT as we evaluated before and after the experiments, see fig. S1). The second experimental set up is based on a high sensitivity AFM inside of a vacuum chamber. Each of them uses a Dulcinea control unit (Nanotec Electronica SL). The images were processed with WSxM software [22]. Cobalt-coated PPP-MFMR NanoSensors cantilevers were used in dynamic mode. The stiffness and resonance frequency of the cantilevers used for the experiments reported in figs. 1 and 2 were k= 1.5 N/m, $f_0$= 69 kHz and k= 1.2 N/m and $f_0$= 59 kHz respectively (k was determined using Sader's method [23]). We have carefully characterized each tip in order to determine their magnetic properties [18, 24]. The magnetic signal is recorded in retrace mode (equivalent to lift mode[TM]): first a topography line is acquired, then using the information obtained in this first scan, a second scan is performed where the topography feedback is disable and the tip follows the topography contour far away from the surface (20-70 nm). The frequency shift images shown in figs. 1 and 2 were acquired at a lift distance of 50 nm with oscillation amplitudes ranging between 4-7 nm. With these amplitudes it is reasonable to use the linear approximation for the force gradient [25, 26]

$$\Delta f = -\frac{f_o}{2k} \cdot \frac{\partial F}{\partial z} \quad (1).$$

Notice that in order to measure weak interactions with small oscillation amplitudes, the best instrumental option is a phase lock loop, that keeps the system at resonance as the tip scans the surface. Using large oscillation amplitudes as in ref. [12] (~100 nm amplitude at 50 nm lift distance according to ref. [27]), the measured signal increases but the interpretation of the experiments becomes very complicated. Moreover, we find the images irreproducible and slight variations in the working conditions change the contrast of the steps in the magnetic signal (as expected under non-linear conditions). This can be readily seen in fig. S2, that shows two consecutive images with slight variations in the imaging conditions. On the contrary, in the linear regime (using low oscillation amplitude) the imaging process is absolutely reproducible as shown in fig. S3, where a series of frequency shift images (obtained in the same region displayed in fig. 1) are measured at very different lift distances.

Fig. 1 portraits the main result using the first experimental set up. Fig. 1a is a edge enhanced AFM topographic image of a ZYH-HOPG surface. The magnetic state of both tip and sample were initially prepared as represented in the inset of fig. 1b using a magnet. Fig. 1b-1f are the corresponding frequency shift signal taken at different magnetic fields. Any long range interaction, such as the tip-sample magnetic force, should be reflected in this magnitude. The insets of these figures represent the tip-sample magnetic states, according to the above discussion, for different $H_E$ values. Fig. 1b, taken in remanence, only shows bright steps and not a bright and dark contrast distribution as expected for a sample with magnetic features oriented in up and down directions. Moreover, in this situation a majority of dark contrast –corresponding to attractive force- is expected. As we vary the external magnetic field (fig. 1b-1f), overcoming first the sample coercive field (fig. 1c $H_E$=+35mT) and then the tip one (Fig. 1d $H_E$=+60 mT), the contrast along the steps, that is the tip-sample magnetic force, remains constant in obvious contradiction with the expected orientation of the magnetizations shown in the insets. A similar situation is observed when the external magnetic field is reversed (fig. 1e-1f). This experiment suggests that the contrast observed along the steps is not of magnetic origin (this is a standard procedure to discard magnetic interaction in AFM images [28]). This conclusion is also valid even assuming that the coercive field for the steps is different from 20 mT. We certainly know that in this experiment we are reversing the tip magnetization and therefore, if the steps exhibit a defined magnetization a contrast inversion in the MFM signal is expected.

As a general rule, measuring AFM signals along steps is always a difficult task. The origin of the contrast observed in graphite with MFM could be a combination of electrostatic interactions (the tips are covered with a metal and therefore they are very sensitive to electrostatic forces) and artifacts: the contrast generated by the topography in the MFM image is not constant at constant z distance and it can cause meaningless contrast along the steps [24]. This is a consequence of the different force gradient near and far away from the surface.

It is well known [29] that steps in conducting surfaces exhibit a dipole that locally changes the surface potential along the steps. On the other hand, it is also known that nucleation preferentially occurs along the steps, therefore we also expect molecule adsorption that can also vary the surface potential. Assuming a defined magnetization along the steps, the long range interactions observed (far away from the surface) by a cobalt coated tip is a combination of both electrostatic and magnetic forces: $F=F_E+F_B$. F being the total long range force observed by the tip, $F_E$ the force produced by the electric dipole along steps and $F_B$ the magnetic force produced by the magnetic dipole along the steps as well. The separation of these interactions can be done by combining KPFM [30, 31] and MFM. KPFM minimizes the electrostatic interaction compensating the local contact potential between tip and sample. To further improve our sensitivity [32] we have carried out experiments which combine both techniques in a high vacuum chamber with a base pressure of $10^{-6}$ mbar (the Q factor at this pressure is 6850, a factor of 49 higher than in air ambient conditions). Fig. 2a is a topographic image of a recently cleaved ZYH-HOPG surface taken at P$\approx 10^{-6}$ mbar using the frequency modulation mode described by [32]. In order to enhance the step edges we are showing the derivative of the topography image (fig. 2b). For reasons still under discussion, graphite exhibits a marked distribution of electrostatic potential on its surface [33] that can be easily measured by KPFM using metallic cantilevers. In our high vacuum experiments we take advantage of the long range nature of both electrostatic and magnetic interactions to prove that within our sensitivity (much higher than in air ambient conditions) we are not able to detect any interaction that can be attributed to magnetic signals. Fig. 2c is a KPFM image taken simultaneously with the topographic image. An advantage of KPFM is that it provides electrostatic images of the sample surface at distances where the van der Waals interactions are still relevant (in this case the KPFM image is taken at the same distance as the topography image). Similar KPFM images can also be easily measured with Pt covered tips, discarding any magnetic component of this signal. Fig. 2d is again a KPFM image of the surface where the tip is lifted 50 nm to avoid short range interactions. The electrostatic signal is basically the same but slightly smoothed by the 50 nm lift distance. Fig. 2e is the frequency shift, simultaneously measured with fig. 2d that gives no signal within our experimental error.

The point to be stressed in this measurement is that we are separating the electrostatic interaction, that goes to the KPFM image (fig. 2d), from the magnetic signal, that should be exclusively present in the frequency shift image (fig. 2e). Since we are not able to measure any significant signal, we conclude that graphite does not exhibit ferromagnetic interaction along the step edges. The clear electrostatic signal measured at lift distance confirms our high sensitivity and discards any artifact due to tip damage.

The magnitude of our noise can be estimated by measuring an ''empty'' image [34]. The root-mean-square (rms) resonance-frequency noise so evaluated is about 0.2 Hz. Using equation (1), this translates in a minimum detectable magnetic signal of ~8 μN/m (in good agreement with ref. [24]). Since fig. 2e does not show any signal along the steps this would be the upper bound for the magnetic force gradient. Moreover, we can take advantage of an instrumental artifact seen in fig. 2e to estimate an even more cautious upper bound for the magnetic force gradient in graphite. While most of the electrostatic signal goes to the KPFM channel, there is still some signal that leaks to the frequency shift image. The origin of this artifact is the KPFM feedback that cannot perfectly compensate the electrostatic signal. This is the conventional error signal of any feedback system and it is also seen when using Pt covered tips, discarding again any magnetic origin. The frequency shift of this region is 0.4 Hz, so we know that the sensitivity of the frequency shift image is at least this one. The corresponding force gradient obtained using equation (1) is 16 μN/m, a factor of 6 lower than the lowest theoretical bound expected (100 μN/m at 50 nm) for the magnetic force gradient in ref. [12]. We consider this a very prudent upper bound for the force gradient produced by any magnetic field on the HOPG surface.

To sum up, the contrast observed along the steps on a graphite surface remains unmodified when an external magnetic field is applied. This indicates a non magnetic origin for the signal observed along the steps. By combining KPFM and MFM we are able to discount the contribution of the electrostatic signal and this allows us to give an upper bound for a magnetic signal along the graphite steps of 16 μN/m.

## Acknowledgements

The authors acknowledge the financial support from the CM grants S2009/MAT-1467and from the Spanish Ministerio de Ciencia e Innovación through the projects, MAT2007-65420-C02-01, MAT2007-66476-C02-01 and MAT2008-02929-NAN. D. Martínez thanks for FPU grant with reference number AP20050079 and M. Jaafar for the JdC contract.

**Figures captions**

**Fig. 1** AFM images in ambient conditions under an external magnetic field a) Edge enhanced topography showing a 2.5μm x 2.5μm ZYH-HOPG typical area with a large density of steps. b-f) MFM images taken at a lift distance of 50 nm using a 5 nm oscillation amplitude. As the external magnetic field ($H_E$) is scanned the contrast along the steps remains unmodified. The insets of figures b-f indicate the expected tip-sample magnetic configuration as $H_E$ is varied.

**Fig. 2** AFM/KPFM/MFM images taken in high vacuum with a cobalt coated probe. a) 3μm x 3μm AFM topography of a ZYH-HOPG surface measured in high vacuum conditions to increase the AFM sensitivity. b) Edge enhanced image of fig. (a) showing the surface steps. c) KPFM image simultaneously taken with (a), showing electrostatic domains and steps on the sample surface (the potential difference between bright and dark areas is 200 mV). d) KPFM image taken in retrace at 50 nm lift distance (this distance is measured respect to the tip-sample distance used during the topography image). The electrostatic distribution seen in (c) is reproduced in (d) but somehow *unfocused* as expect for this lift distance. e) Frequency shift image taken simultaneously with (d) Since the KPFM technique compensates the electrostatic interaction the remaining signal should correspond to other possible long range interactions. The image does not show any significant contrast suggesting absence of magnetic interaction within our sensitivity (enhanced by the high Q obtained in vacuum). The origin of the light shadows is instrumental: the KPFM feedback does not perfectly compensate the electrostatic interaction leaking a small part of it to the frequency shift image. We take advantage of this issue to estimate a conservative upper bound for the magnetic force gradient of 16 μN/m. The total frequency shift variation in figure (e) is 0.4 Hz.

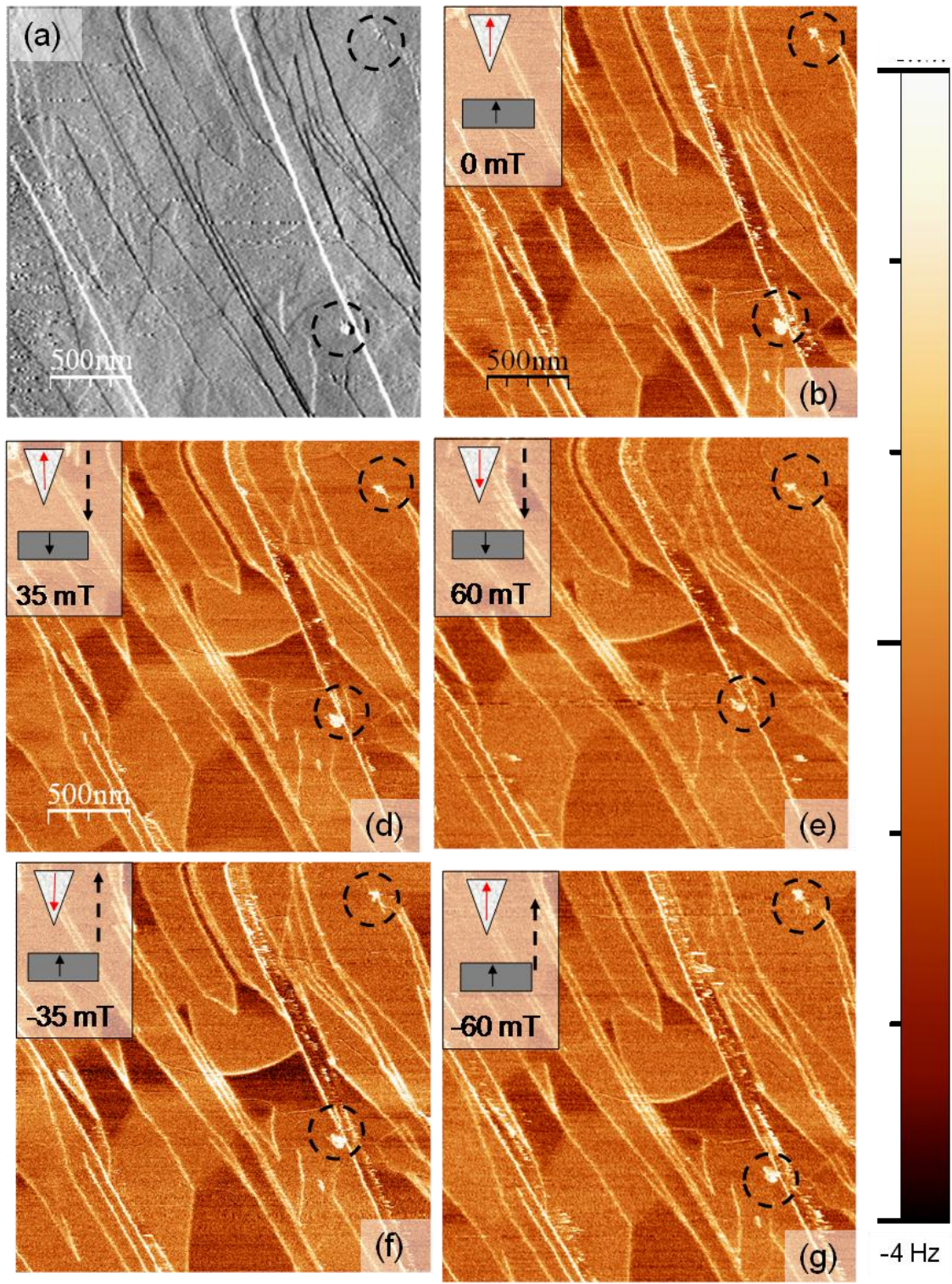

Fig. 1

Fig. 2

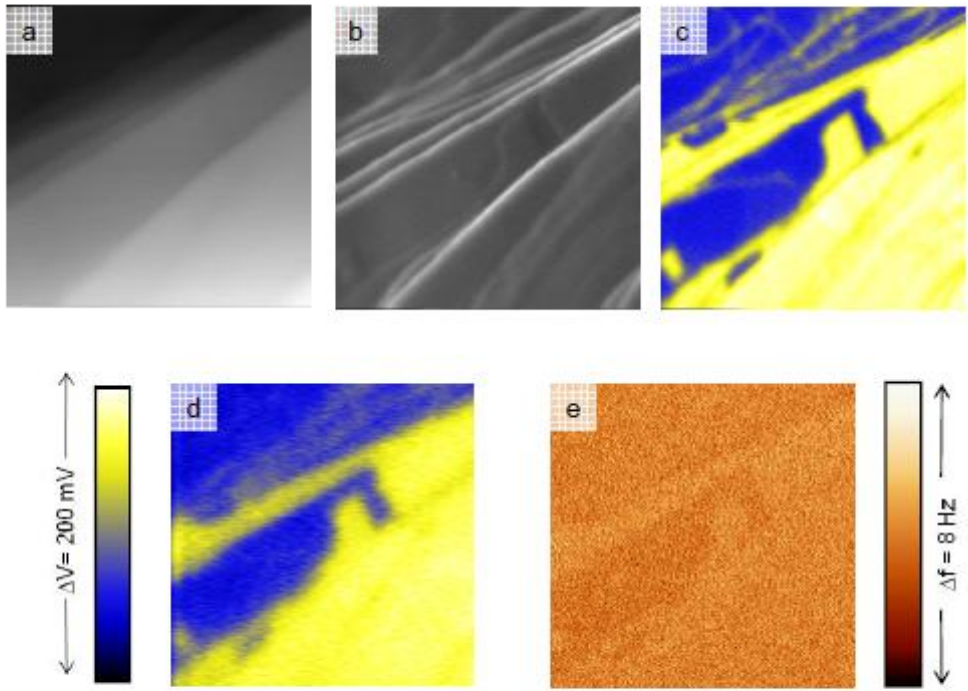

# Supplementary information for

An upper bound for the magnetic force gradient in graphite

David Martínez-Martín[1], Miriam Jaafar[1], Rubén Pérez[2], Julio Gómez-Herrero[1] and Agustina Asenjo[2].

[1]Dpto. Física de la Materia Condensada, UAM, 28049 Madrid, Spain

[2]Dpto. Física Teórica de la Materia Condensada, UAM, 28049 Madrid, Spain

[3]Instituto de Ciencia de Materiales de Madrid, CSIC, 28049 Madrid, Spain

**Tip characterization.**

The magnetic probes used in our experiments have been characterized by scanning the magnetic field over a reference sample (a magnetic hard drive) as shown in figs. S1a-b. The hysteresis loop can be obtained from this 3D mode images [1] (a more detailed explanation can be found in ref. [2]). Fig S1c shows a typical hysteresis loop obtained in that way for one of the magnetic tips used in our experiments. Notice that after saturating the sample in opposite directions, the MFM contrast is completely reversed (see Fig S1 d-g).

**Large oscillation amplitudes**

The signal to noise ratio can be increased by using large oscillation amplitudes. The problem of AFM images with large amplitude oscillations is that it is very easy to mix van der Waals forces (medium range interactions) with electrostatic and magnetic forces (long range interactions). In addition, the simple analysis in terms of linear theory becomes meaningless [3, 4]. Imaging at 50 nm lift distance with low oscillation amplitudes ensures that you are only really measuring long range interactions. More precisely, the expression

$$\Delta \Phi = -\frac{Q}{k} \cdot \frac{\partial F}{\partial z}$$

is just valid for low amplitude oscillations. For large amplitude oscillations the system becomes highly non linear and this expression it is not valid any longer. The phase shift becomes much more complicated and must be calculated as a convolution of a semispherical weight function with the tip-sample interaction [4]. In order to measure weak interactions with small oscillation amplitudes the best instrumental option is a phase lock loop that keeps the system at resonance as the tip scans the surface.

For the sake of comparison, we have carried out experiments using large oscillation amplitudes, simulating the operating conditions in ref.[5]. We find the images completely irreproducible and small variations in the imaging conditions change the contrast of the steps in the frequency/phase shift signal (as expected under non linear conditions). This can be readily seen in Figure S2 that shows two consecutive images with small variations in the imaging conditions. One may argue that we have not carried out the experiment carefully enough but the real problem is that under these conditions the theory anticipates irreproducible results. On the contrary low amplitude oscillations produce perfectly reproducible results as shown in figure S3.

Figures.

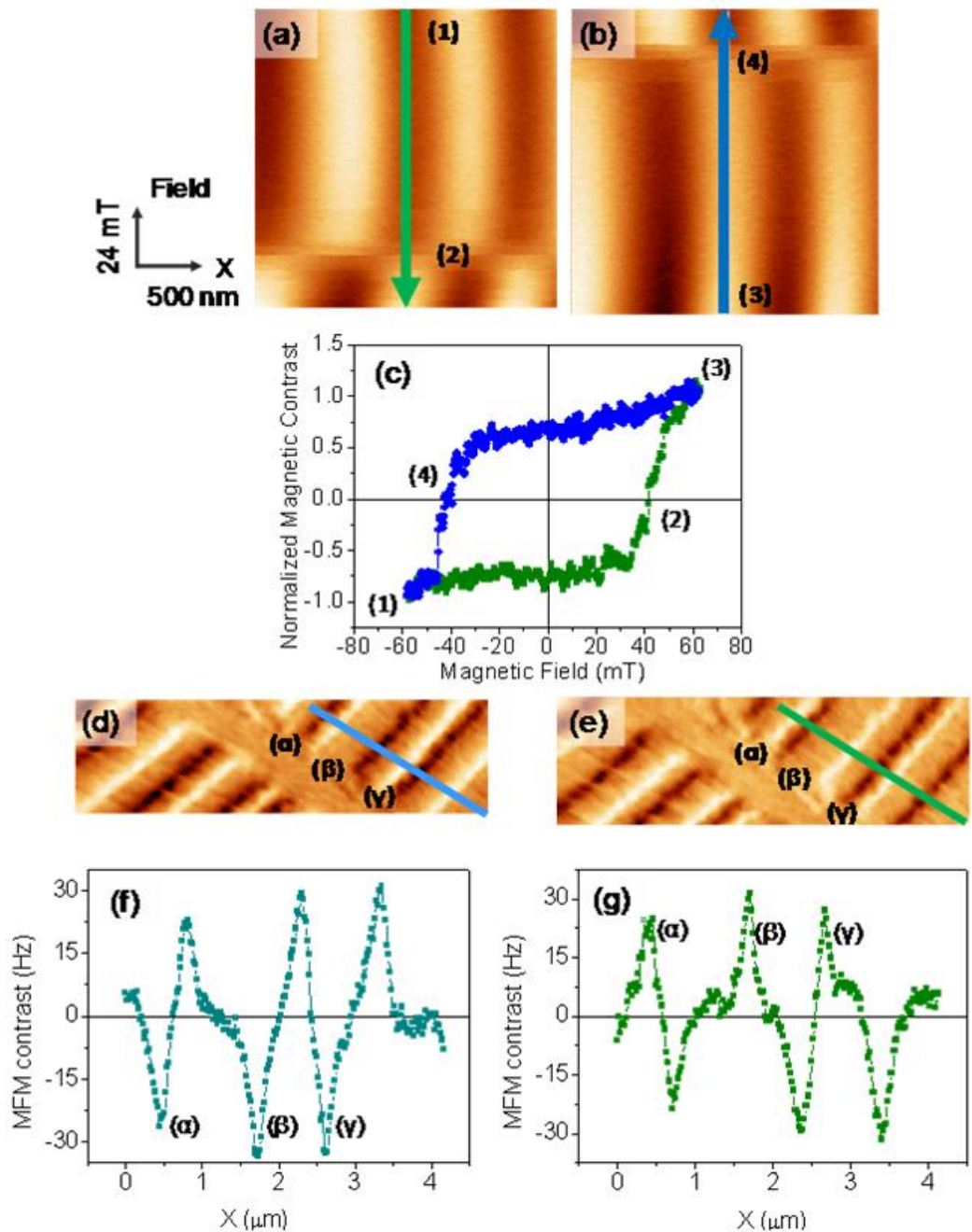

**FigureS1-** (a) and (b) are frequency shift images as a function of the tip position (along one scan line in a reference sample) and the applied magnetic field (vertical direction), the field varies continuously along the arrows directions according to hysteresis loop ;(c) hysteresis loop of the MFM probe obtained along the vertical lines in (a) and (b). (d) and (e) are MFM images of a reference sample after applying +60 mT and -60 mT respectively. Notice that the contrast is complete reversed as we can observed in the corresponding profiles(f) and (g) The coercive field of the sample is around 200 mT so all the changes in the magnetic contrast are due to the switching of the magnetization of the MFM probe.

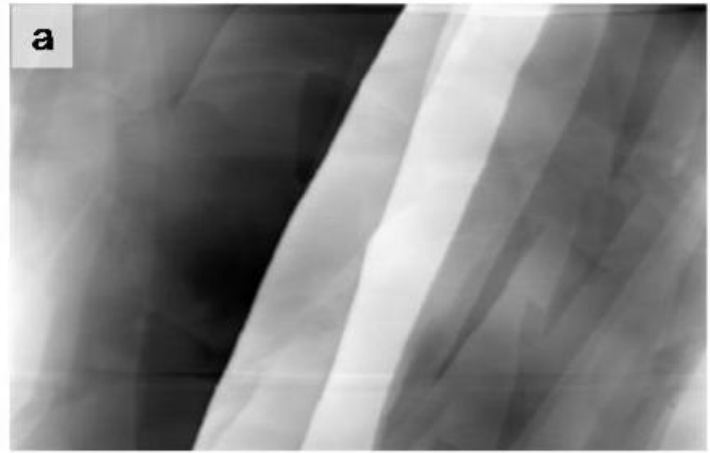

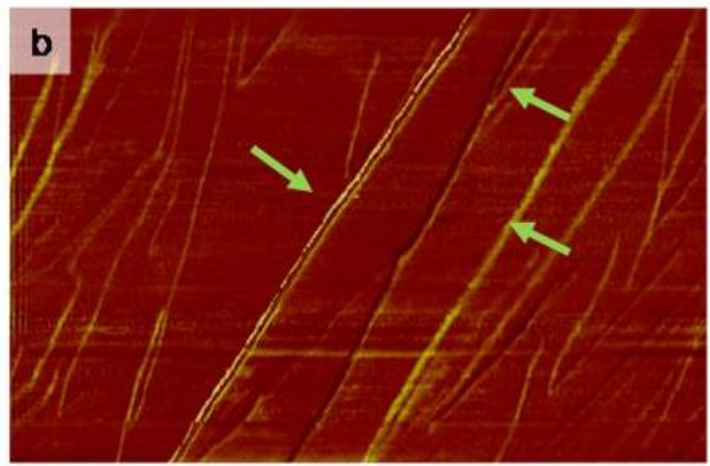

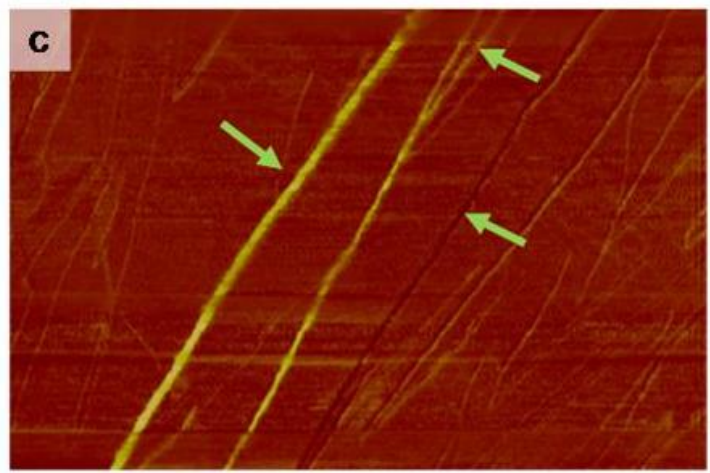

**Figure S2 -** Large amplitude phase shift images: a) HOPG topography. (b) and (c) are two images of the phase shift at a lift distance of 50 nm with small variation in the imaging condition. The contrast along the step has changed dramatically without any applied magnetic field or changing the tip magnetization. Image size: 3.5 μm x 2.8 μm.

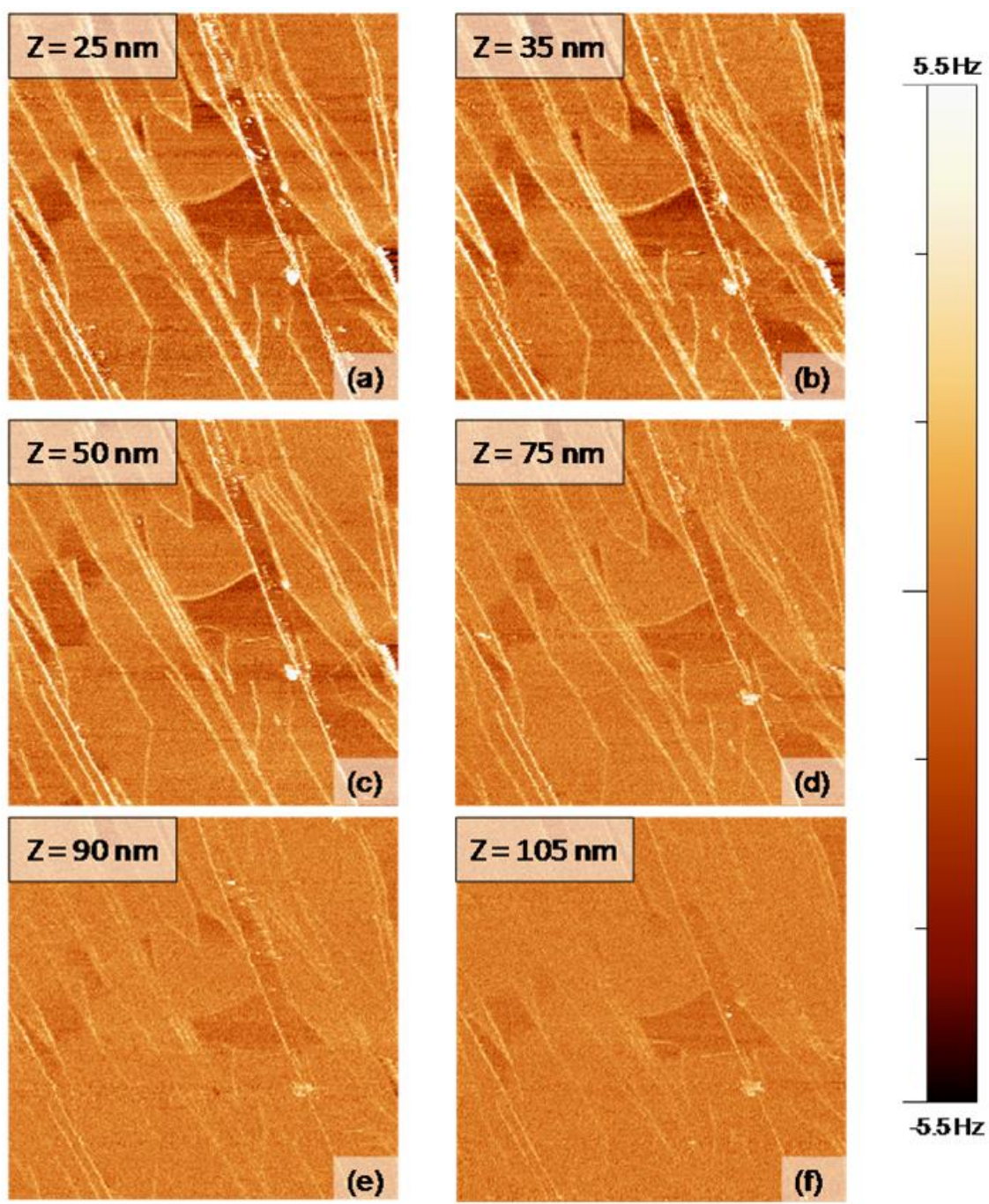

**Figure S3** - MFM images obtained with low amplitude of oscillation at different lift distances. Image size: 2.4 µm x 2.4 µm .